\documentclass[preprint,5p, twocolumn]{elsarticle}

\usepackage{lineno,hyperref}
\usepackage{lineno}
\usepackage{graphicx}
\usepackage{subcaption}
\usepackage{amsmath}
\usepackage{verbatim}
\usepackage{amssymb}
\usepackage{color}

\usepackage{mathrsfs}
\modulolinenumbers[5]

\journal{Journal of \LaTeX\ Templates}

\begin{document}

\begin{frontmatter}

\title{Kink-antikink collision in a Lorentz-violating $\phi^4$ model}

\author{Haobo Yan}
\author[mysecondaryaddress]{Yuan Zhong\corref{mycorrespondingauthor}}
\cortext[mycorrespondingauthor]{Corresponding author}
\ead{zhongy@mail.xjtu.edu.cn}
\address{School of Science, Xi'an Jiaotong University, \\
No. 28 West Xianning Road, Xi'an 710049, People's Republic of China}
\author{Yu-Xiao Liu}
\address{Institute of Theoretical Physics \& Research Center of Gravitation, Lanzhou University,\\
No. 222 South Tianshui Road, Lanzhou 730000, People's Republic of China}
\author{Kei-ichi Maeda}
\address{Department of Physics, Waseda University, Shinjuku, Tokyo 169-8555, Japan}
\address{Waseda Institute for Advanced Study {\rm (WIAS)},
Waseda University, Shinjuku, Tokyo 169-8050, Japan}

\begin{abstract}
In this work, kink-antikink collision in a two-dimensional Lorentz-violating $\phi^4$ model is considered. It is shown that the Lorentz-violating term in the proposed model does not affect the structure of the linear perturbation spectrum of the standard $\phi^4$ model, and thus there exists only one vibrational mode. The Lorentz-violating term impacts, however, the frequency and spatial wave function of the vibrational mode. As a consequence, the well-known results on $\phi^4$ kink-antikink collision will also change. Collisions of kink-antikink pairs with different values of initial velocities and Lorentz-violating parameters are simulated using the Fourier spectral method. Our results indicate that models with larger Lorentz-violating parameters would have smaller critical velocities $v_c$ and smaller widths of bounce windows. Interesting fractal structures existing in the curves of maximal energy densities of the scalar field are also found.
\end{abstract}


\end{frontmatter}


\section{Introduction}

The domain wall is a simple type of topological soliton that exists in many nonlinear scalar field models. It plays an important role in many branches of physics. For instance, the duality between the sine-Gordon model and massive Thirring model provides the simplest example of bosonization in condensed-matter physics~\cite{Coleman1985}. Cosmic domain walls can exist if the Universe was proceeded by some first-order phase transitions~\cite{VilenkinShellard2000,Vachaspati2006}. It is even proposed that all of us might live on a four-dimensional domain wall embedded in a five-dimensional space-time~\cite{RubakovShaposhnikov1983,Akama1982,Arkani-HamedDimopoulosDvali1998a,RandallSundrum1999a,RandallSundrum1999,ShiromizuMaedaSasaki2000}; also see \cite{DzhunushalievFolomeevMinamitsuji2010,Liu2018} for reviews. A domain wall in $(1+1)$ dimensions is also called a kink.

 The collision between non-integrable kinks is an important topic in the study of kinks. For the simplest kink-antikink collision, it is convenient to take the velocities of the kink and antikink as $v_0$ and $-v_0$, respectively. Such collisions are referred to as velocity-symmetric collisions or symmetric collisions in this paper. All the kink-antikink collisions mentioned below are of this type, only the velocity of the kink $v_0$ will be specified.

 In integrable models, e.g., the sine-Gordon model, a kink and an antikink simply pass through each other after the collision, while in non-integrable models, however, the outcome of a kink-antikink collision sensitively depends on the initial velocity of the kink $v_0$. Taking the $\phi^4$ model as an example, if $v_0$ is larger than the critical velocity $v_c\approx 0.26$, one observes inelastic scatterings with the emission of scalar radiation~\cite{Sugiyama1979}. If $v_0<v_c$, one usually observes a spatially localized oscillating structure called a bion or oscillon, which is a bound state of the kink and antikink~\cite{Kudryavtsev1975}. However, if $v_0$ lies in some narrow intervals below $v_c$, one would observe the interesting $n$-bounce phenomenon; that is, after colliding $n$ times, kinks escape rather than trapping into a bion~\cite{Moshir1981,CampbellSchonfeldWingate1983}. These magical intervals are called $n$-bounce windows ($n$BWs), and have been found in many non-integrable models~\cite{PeyrardCampbell1983,CampbellPeyrardSodano1986}. More interestingly, all the $n$-bounce windows form a fractal structure, which means that one may find some $(n+1)$-bounce windows by zooming into the boundaries of an $n$-bounce window~\cite{AnninosOliveiraMatzner1991}.

According to the widely accepted Campbell-Schonfeld-Wingate (CSW) mechanism~\cite{CampbellSchonfeldWingate1983}, the bounce-window phenomenon is caused by a resonant energy exchange between the vibrational mode and zero mode around a kink (antikink). However, some recent works reveal that the CSW mechanism is insufficient to describe bounce phenomena found in some non-integrable models in which the kink either has no vibrational mode at all~\cite{DoreyMershRomanczukiewiczShnir2011}, or only has quasinormal modes~\cite{DoreyRomanczukiewicz2018}. Therefore, by exploring kink-antikink collisions in various kinds of models one may find new phenomena with new physics.

Beyond the $\phi^4$ model, there are many works discussing kink-antikink collisions in models with both polynomial potentials~\cite{GaniKudryavtsevLizunova2014,GaniLenskyLizunova2015,BelendryasovaGani2019,ChristovDeckerDemirkayaGaniEtAl2018,ChristovDeckerDemirkayaGaniEtAl2019,GomesSimasNobregaAvelino2018,MendoncaOliveira2015,MendoncaOliveira2015a,MendoncaDeOliveira2019,AdamOlesRomanczukiewiczWereszczynski2019,AdamOlesRomanczukiewiczWereszczynski2019a,AdamOlesRomanczukiewiczWereszczynski2019b,ChristovDeckerDemirkayaGaniEtAl2020} and triangular potentials~\cite{GaniKudryavtsev1999,GaniMarjanehAskariBelendryasovaEtAl2018,BazeiaBelendryasovaGani2018,BazeiaGomesNobregaSimas2019,BazeiaGomesNobregaSimas2019a}, with noncanonical dynamics~\cite{GomesMenezesNobregaSimas2014,ZhongDuJiangLiuEtAl2019}, and with multiple scalar components~\cite{Alonso-Izquierdo2018a,Alonso-Izquierdo2018b,Alonso-IzquierdoBalseyroSebastianGonzalezLeon2018,Alonso-Izquierdo2019,Alonso-Izquierdo2019b}. In addition, multi-kink collisions have also been extensively studied recently~\cite{MarjanehGaniSaadatmandDmitrievEtAl2017,MarjanehSaadatmandZhouDmitrievEtAl2017,MarjanehAskariSaadatmandDmitriev2018,GaniMarjanehSaadatmand2019}.
All these works assume Lorentz invariance of their models.

However, as pointed out in Ref. \cite{AllenYokoo2004}, every current candidate for a superunified theory\footnote{A “superunified theory” is one which includes all known physical phenomena, and is valid up to Planck's energy.} contains some potentials for Lorentz violation, and the same is true for more restricted theories that attempt to treat quantum gravity alone. Theories with the potential for Lorentz violation, including superstring/M/brane theories, canonical and loop quantum gravity, non-commutative spacetime geometry, non-trivial space-time topology, and so on; see \cite{CarrollFieldJackiw1990,CarrollHarveyKosteleckyLaneEtAl2001,KosteleckyLehnert2001,BergerKostelecky2002,KosteleckyMewes2002,XiaoMa2009a} for part of the original works, and \cite{Mattingly2005,XiaoMa2009,Bietenholz2011,Liberati2013,KosteleckyRussell2011,Torri2020} for recent reviews. Thus, it is interesting to search for kink solutions in Lorentz-violating scalar field theories, and study how Lorentz violation impacts the properties of the kinks and their collision.

Analytical static and traveling kink solutions have been found in some Lorentz-violating scalar field models with single or multi-field components~\cite{BazeiaMenezes2006,BazeiaFerreiraGomesMenezes2010,SouzaDutraCorrea2011,PassosAlmeidaBritoMenezesEtAl2018}, and these solutions have been applied in many related issues, e.g., entropic information~\cite{CorreaRochaSouzaDutra2015}, the Kondo effect~\cite{BazeiaBritoMota-Silva2016}, and trapping fermions~\cite{CorreaPaulaSouzaDutraFrederico2018}.

In this paper, kink-antikink interaction is considered by using the traveling kink solution reported in Ref.~\cite{BazeiaMenezes2006}. In the next section, the single-field Lorentz-violating model as well as the corresponding kink solution is reviewed. After an analysis of the linear stability, a numerical simulation of the kink-antikink collision is conducted.

\section{Model and solution}
The $(1+1)$-dimensional Lorentz-violating scalar field model of \cite{BazeiaMenezes2006} takes the following Lagrangian density:
\begin{eqnarray}
\mathcal{L}=\frac{1}{2} \eta^{\mu \nu}\partial_{\mu} \phi \partial_{\nu} \phi+\frac{1}{2} \kappa^{\mu \nu} \partial_{\mu} \phi \partial_{\nu} \phi-V(\phi),
\label{lagrange}
\end{eqnarray}
with $\eta^{\mu \nu}\equiv \begin{pmatrix} 1 & 0 \\ 0 & -1 \end{pmatrix}$ and $\kappa^{\mu \nu}\equiv \begin{pmatrix} 0 & \alpha \\ \alpha & 0 \end{pmatrix}.$
The equation of motion reads
\begin{equation}
\label{eomMoving}
\frac{\partial^{2} \phi}{\partial t^{2}}-\frac{\partial^{2} \phi}{\partial x^{2}}+2 \alpha \frac{\partial^{2} \phi}{\partial x \partial t}+V_\phi=0,
\end{equation}
with $V_\phi\equiv \frac{d V}{d\phi}$.
The violation of Lorentz symmetry is described by the parameter $\alpha$, which is assumed to be non-negative, $\alpha\geq 0$, in this paper\footnote{The model with $\alpha<0$ can be obtained by applying the time-reversal transformation. Although the theory is not time symmetric and the behaviors are different from the case with $\alpha>0$, the same results are obtained when the space is reversed at the same time, because PT is conserved.}.

\begin{figure}[h]
\centering
\includegraphics[width=0.48\textwidth]{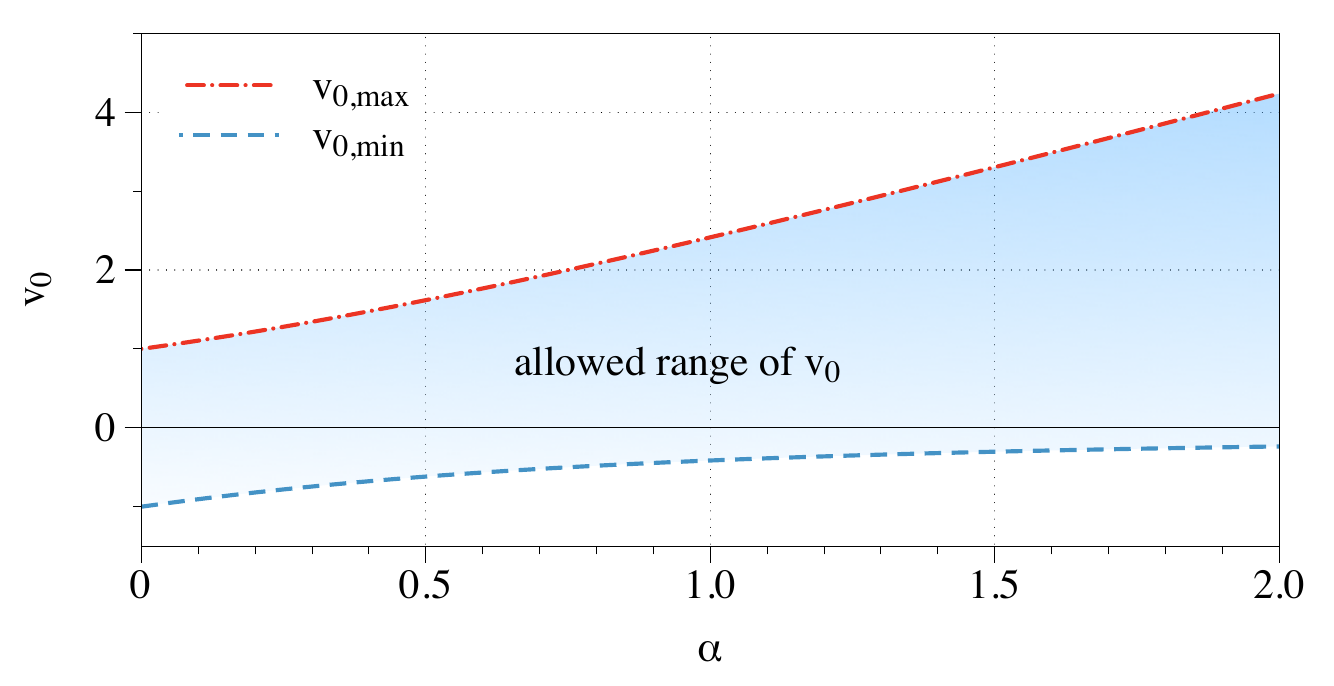}
\caption{Allowed range of $v_0$ for $\alpha\in[0,2]$.}
\label{range}
\end{figure}

Obviously, the Lorentz-violating term does not alter the form of a static solution.
Therefore, it is trivial to find a static kink solution. For example, a standard static kink solution
\begin{equation}
\phi_s(x)=\tanh x
\label{before_boost}
\end{equation}
is obtained by taking the potential as
\begin{equation}
V(\phi)=\frac{1}{2}(1-\phi^{2})^{2}.
\label{Vphi}
\end{equation}
The real challenge is to find a traveling kink solution. The breaking of Lorentz invariance makes it difficult, in general, to find a boost transformation that helps derive traveling solutions from static ones. Fortunately, as mentioned in Ref.~\cite{BazeiaMenezes2006}, the present model is invariant under a deformed boost transformation $x\to x'=\gamma(x-vt)$, where the deformed Lorentz factor is given by $\gamma\equiv1 / \sqrt{1-v^{2}+2 \alpha v}$ in the natural units\footnote{The model is also translation invariant, so it possesses a deformed Poincar\'e symmetry.}.  Therefore, the traveling kink and antikink solutions take the following forms:
\begin{eqnarray}
\phi_K(x_0, v_0)&=&\tanh (\gamma(x-x_0-v_0 t)),\\
\phi_{\bar K}(x_0, v_0)&=&-\phi_K(x_0, v_0),
\label{boost}
\end{eqnarray}
where $x_0$ and $v_0$ are the initial position and velocity of the kink/antikink, respectively. A positive (negative) $v_0$ means a kink/antikink moving to the right (left).
With the definition of $\gamma$, the allowed range of initial velocity varies with the value of the parameter $\alpha$:
\begin{equation}
v_{0,\text{mim}}=-\sqrt{1+\alpha^{2}}+\alpha < v_0 < \sqrt{1+\alpha^{2}}+\alpha= v_{0,\text{max}},
\end{equation}
which is shown in Fig.~\ref{range}.

\begin{figure}[h]
\centering
\includegraphics[width=0.48\textwidth]{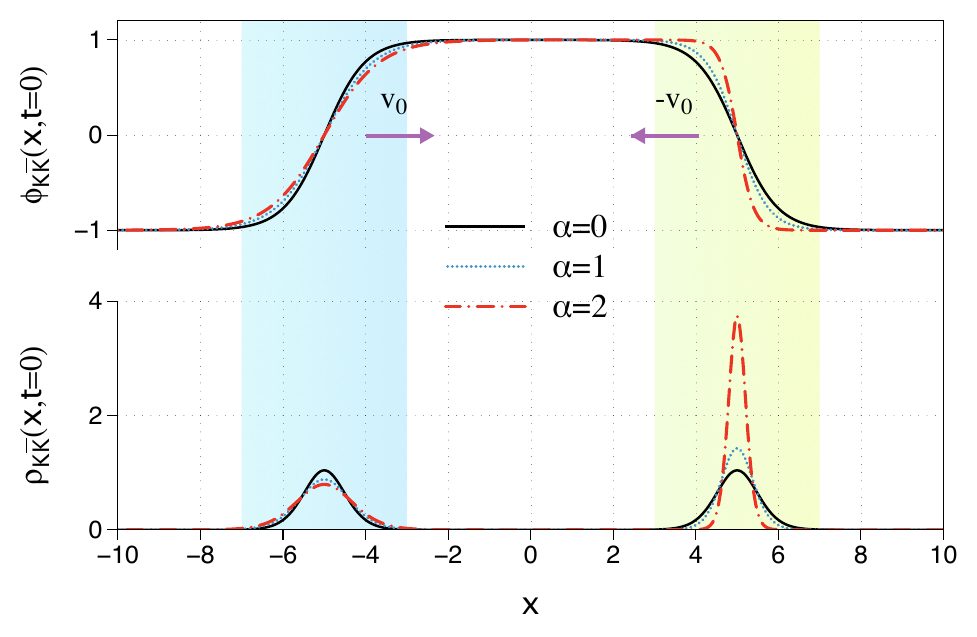}
\caption{Configuration of $\phi_{K\bar{K}}$ and its energy density at time $t=0$ for  $\alpha=0,1,2$. $x_0=5$ and $v_0=0.2$ have been taken. Obviously, for $\alpha>0$, an antikink moving to the left has smaller width and larger energy density than a kink moving to the right with the same speed.}
\label{phiandRho}
\end{figure}

The energy density for a traveling solution is \cite{BazeiaMenezes2006}
\begin{equation}
\rho(x,t)=\frac{1}{2}\left(\frac{\partial \phi}{\partial t}\right)^{2}+\frac{1}{2}\left(\frac{\partial \phi}{\partial x}\right)^{2}+V(\phi).
\label{rhos}
\end{equation}
Obviously, in the Lorentz-violating model, the width and energy density of a moving kink depend on both the magnitude and direction of its velocity because of breaking of time symmetry. To see this phenomenon clearly, consider the linear superposition of a pair of a moving kink and antikink:
\begin{eqnarray}
\phi_{K\bar{K}}(x,t)=\phi_K(-x_0, v_0)+\phi_{\bar K}(x_0, -v_0)-1.
\label{eqInitial}
\end{eqnarray}
In Fig.~\ref{phiandRho}, the configuration of $\phi_{K\bar{K}}$ and its energy density at time $t=0$ is plotted for $\alpha=0,1,2$. The asymmetry between the right-moving kink and left-moving antikink appears as $\alpha$ increases. An antikink moving to the left has smaller width and larger energy density than those of a kink moving to the right with the same speed for $\alpha>0$. It is the opposite for $\alpha<0$.
\section{Linear stability and vibrational mode}
\label{secLinear}
Before the discussion of kink-antikink collision, it is important to study the linear perturbation of the static kink solution. First, a good kink solution should be stable against the linear perturbation. In addition, according to the CSW mechanism, the existence of vibrational modes in a linear spectrum is closely connected with the bounce-window phenomenon.

To derive the equation of motion for a small perturbation $\delta \phi (t,x)\ll 1$ vibrating around the static kink background $\phi_s(x)$, the action is expanded up to the second order of $\delta \phi$:
\begin{equation}
S=S^{(0)}+\delta^{(1)} S+\delta^{(2)}  S +\mathcal{O}(\delta \phi^3),
\end{equation}
where ${\displaystyle \delta^{(2)} {S}= \int \delta^{(2)} \mathcal{L} \, {d}^2 x}$ with
\begin{equation}
\delta^{(2)} \mathcal{L}=\frac{1}{2} \partial_{\mu} \delta \phi \partial^{\mu} \delta \phi+\frac{1}{2} \kappa^{\mu \nu} \partial_{\mu} \delta \phi \partial_{\nu} \delta \phi-\frac{1}{2} V_{\phi\phi} \delta \phi ^2.
\end{equation}
After taking the variation of $\delta^{(2)} {S}$ with respect to $\delta\phi$, the linear perturbation equation is obtained:
\begin{eqnarray}
\label{eqPertTensor}
 \partial^{\mu}\partial_{\mu} \delta \phi
+ \kappa^{\mu \nu} \partial_{\mu}\partial_{\nu} \delta \phi
+ V_{\phi\phi} \delta \phi=0.
\end{eqnarray}
The following mode expansion is then introduced:
\begin{equation}
\delta \phi(x,t)=\sum_{n=0}^{\infty} f_{n}(x)\mathrm{e}^{{i} w_n( \alpha x+t)},
\label{series}
\end{equation}
and substituted into Eq.~\eqref{eqPertTensor} to obtain a Schr\"odinger-type equation for $f_{n}(x)$:
\begin{equation}
\label{eqlinearOne}
-\frac{d^2f_n}{dx^2}+V_{\phi \phi } f_n=\tilde{w}_n^2 f_n,
\end{equation}
where $ \tilde{w}_n^2\equiv \left(1+\alpha ^2\right) w_n^2$. It is known from Eq.~\eqref{eomMoving} that a static solution satisfies $\partial_x^2\phi_s=V_\phi$, which means $V_{\phi\phi}=\frac{\partial_x^3\phi_s}{\partial_x\phi_s}$, and Eq.~\eqref{eqlinearOne} can be rewritten as
\begin{equation}
\mathscr{H} f_n=\tilde{w}_n^2 f_n\,,
\end{equation}
where $\mathscr{H}\equiv -\frac {{d}^2}{{d}x^2}+\frac{\ddot \theta}{\theta}$ (with $\theta=\partial_x\phi_s$) is
the Hamiltonian operator.
It can be further factorized as \cite{ZhongGuoFuLiu2018,ZhongLiu2014}
\begin{eqnarray}
\mathscr{H}&=&\mathscr{A} \mathscr{A}^{\dagger}=\left(\frac{{d}}{{d} x}+\frac{\dot{\theta}}{\theta}\right) \left(-\frac{{d}}{{d} x}+\frac{\dot{\theta}}{\theta}\right)\,.
\end{eqnarray}

According to supersymmetric quantum mechanics, the eigenvalues of a system with a factorizable Hamiltonian are always non-negative, i.e., $\tilde{w}_n^2\geq0$, which also means ${w}_n^2\geq 0$. Therefore, the static kink solution $\phi_s(x)$ is stable against linear perturbation.

\begin{figure}[htbp]
\centering
\includegraphics[width=0.49\textwidth]{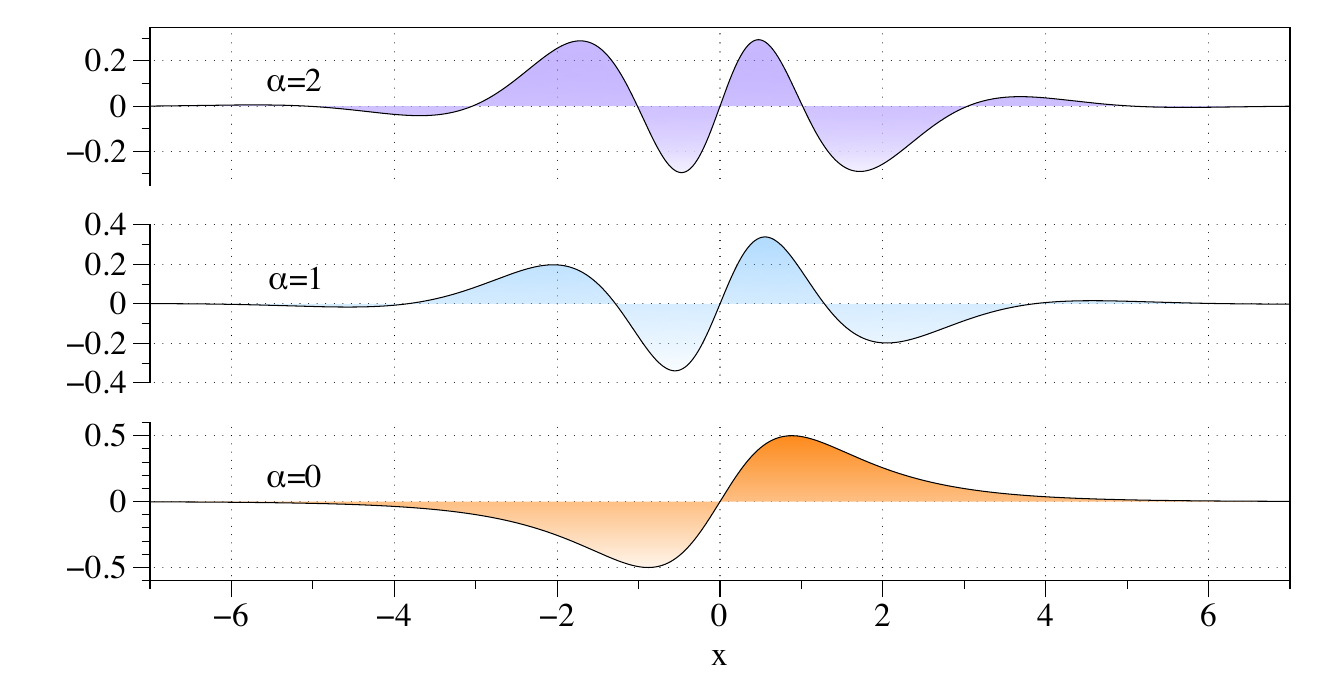}
\caption{Spatial wave function of vibrational mode $g_1(x)\equiv f_1(x)e^{i  \omega_1 \alpha x}$ with $\omega_1=\sqrt{3/(1+\alpha^2)}$ and $\alpha=0,1,2$.}
\label{fig_waveFunction}
\end{figure}

As can be seen from Eq.~\eqref{eqlinearOne}, the Lorentz-violating term does not affect the structure of the linear spectrum of the standard $\phi^4$ model. Thus, for the Lorentz-violating $\phi^4$ model considered here, there are two bound states: the translational mode (zero mode) and vibrational mode, and the corresponding eigenvalues and wave functions are \cite{Vachaspati2006}
\begin{eqnarray}
&&\tilde\omega_0=0,\quad f_0\propto \text{sech}^2 (x),\\
&&\tilde\omega_1^2=3,\quad f_1\propto \text{sech}(x)  \tanh(x).
\end{eqnarray}
However, one should note that the complete spatial wave function for the $n$th mode is $g_n(x)\equiv f_{n}(x)\mathrm{e}^{{i} w_n\alpha x}$, which varies with $\alpha$, except for the zero mode; $\omega_0=0$ and $g_0(x)=f_0(x)$.
Therefore, the Lorentz-violating term does affect the shape of the vibrational mode, as shown in Fig.~\ref{fig_waveFunction}.

Since the vibrational mode is closely related to the bounce-window phenomenon, it is
expected that the results of kink-antikink collision will also change with $\alpha$.
This issue is discussed next.

\section{Simulation and results}
\label{sec_col}
Unlike many integrable models, e.g., the sine-Gorden model, where analytical solutions for multiple kinks can be derived by using methods like B\"acklund transformation, it is extremely difficult to find multi-kink solutions in a non-integrable model. Therefore, to study the collision of a kink-antikink pair in the model proposed herein, one must resort to numerical simulation.

First, the initial condition of the system is taken as
\begin{equation}
\begin{cases}
\phi(x,0)=\phi_{K\bar K}(x,t)|_{t=0},\\
\dot{\phi}(x,0)=\dot{\phi}_{K\bar K}(x,t)|_{t=0},
\end{cases}
\end{equation}
where $\phi_{K\bar K}(x,t)$ defined in Eq.~\eqref{eqInitial} is a superposition of a kink initially at $-x_0$ with velocity $v_0$ and an antikink initially at $x_0$ with velocity $-v_0$.

The dynamical equation \eqref{eomMoving} is solved by using the Fourier spectral method described in Refs.~\cite{Trefethen2001,WangLiuCaiTakeuchiEtAl2012,ZhongDuJiangLiuEtAl2019}. A reasonable numerical solution should satisfy the energy-conservation law, and this fact is used here to test the viability of the proposed numerical solution.

The total energy of the Lorentz-violating field is given by
\begin{equation}
E(t)=\frac{1}{2} \int_{-\infty}^{\infty}\left[\left(\frac{\partial \phi}{\partial t}\right)^{2}+\left(\frac{\partial \phi}{\partial x}\right)^{2}+\left(1-\phi^{2}\right)^{2} \right] \mathrm{d} x.
\label{total_energy}
\end{equation}
For a well-separated kink-antikink pair, the total energy predicted by theory is
\begin{equation}
E_{\rm th}[\phi_{K\bar{K}}]\approx E(v_0)+E(-v_0),
\end{equation}
where $E(v_0)=\gamma(1+\alpha v_0)E_s$ is the energy of a soliton moving with speed $v_0$, and
$
{\displaystyle E_s=\int [(\partial_x\phi)^2/2+V]dx={4\over 3}}
$
is the total energy of the static soliton $\phi_s(x)$.

The conservation of the total energy is checked by evaluating the relative energy error:
\begin{equation}
\delta E\equiv\frac{E_{\rm th}-E_{\rm num}}{E_{\rm th}}.
\end{equation}
Here, $E_{\rm num}$ is the numerical result of the total energy, which is obtained by inserting the numerical solution of $\phi(x,t)$ into Eq. \eqref{total_energy}. In this work, our simulations are implemented with a spatial grid step $\Delta x = 0.2$. The time step $\Delta t$ is automatically determined by the ode45 solver in MatLab (MathWorks, USA). The precision of the calculation can be controled by tuning $\Delta x$ or/and the tolerance options of the ode45 function, i.e., AbsTol and RelTol (see Ref.~\cite{ZhongDuJiangLiuEtAl2019} for details). In our simulation, the tolerance options have been set to ensure that our numerical solutions of $\phi(x,t)$ satisfy $|\delta E|\lesssim10^{-8}$.

\subsection{Impacts on fractal structure}

\begin{figure}[h]
\centering
\includegraphics[width=0.49\textwidth]{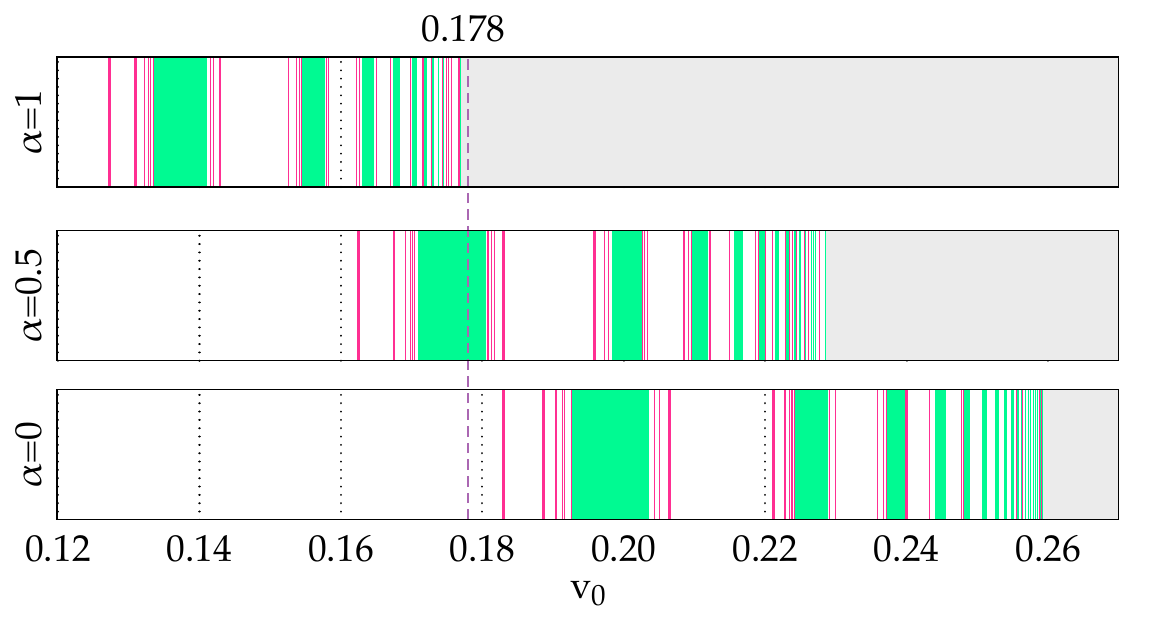}
\caption{Fractal structures of kink-antikink collisions for $\alpha=0, 0.5$ and $1$. The 2BWs 3BWs and inelastic scattering zones are highlighted in green, pink, and gray, respectively. Un-highlighted zones correspond to higher-order bounce windows and bions. Obviously, as $\alpha$ increases, the critical velocities as well as the widths of each 2BW decrease.}
\label{structure}
\end{figure}

To get a global idea of the influence of the Lorentz-violating term, first consider how the fractal structure  would change under different values of $\alpha$. In Fig.~\ref{structure}, the fractal structures (up to three-bounce windows) are plotted for $\alpha=0, 0.5$, and 1. The two-bounce windows (2BWs), three-bounce windows (3BWs), and the inelastic scattering zones are highlighted in green, pink, and gray, respectively. The un-highlighted zones correspond to higher-order bounce windows as well as bions. The critical velocity (the left-hand boundary of the gray zone) decreases from $v_c(\alpha =0)\approx 0.26$, to $v_c(\alpha =0.5)\approx 0.23$, and eventually to $v_c(\alpha =1)\approx 0.177$. As $\alpha$ increases, all the 2BWs (the green zones) move to the left, and their widths decrease as well.

\begin{figure*}[h]
\centering
\begin{subfigure}[b]{0.325\textwidth}
\centering
\includegraphics[width=\textwidth]{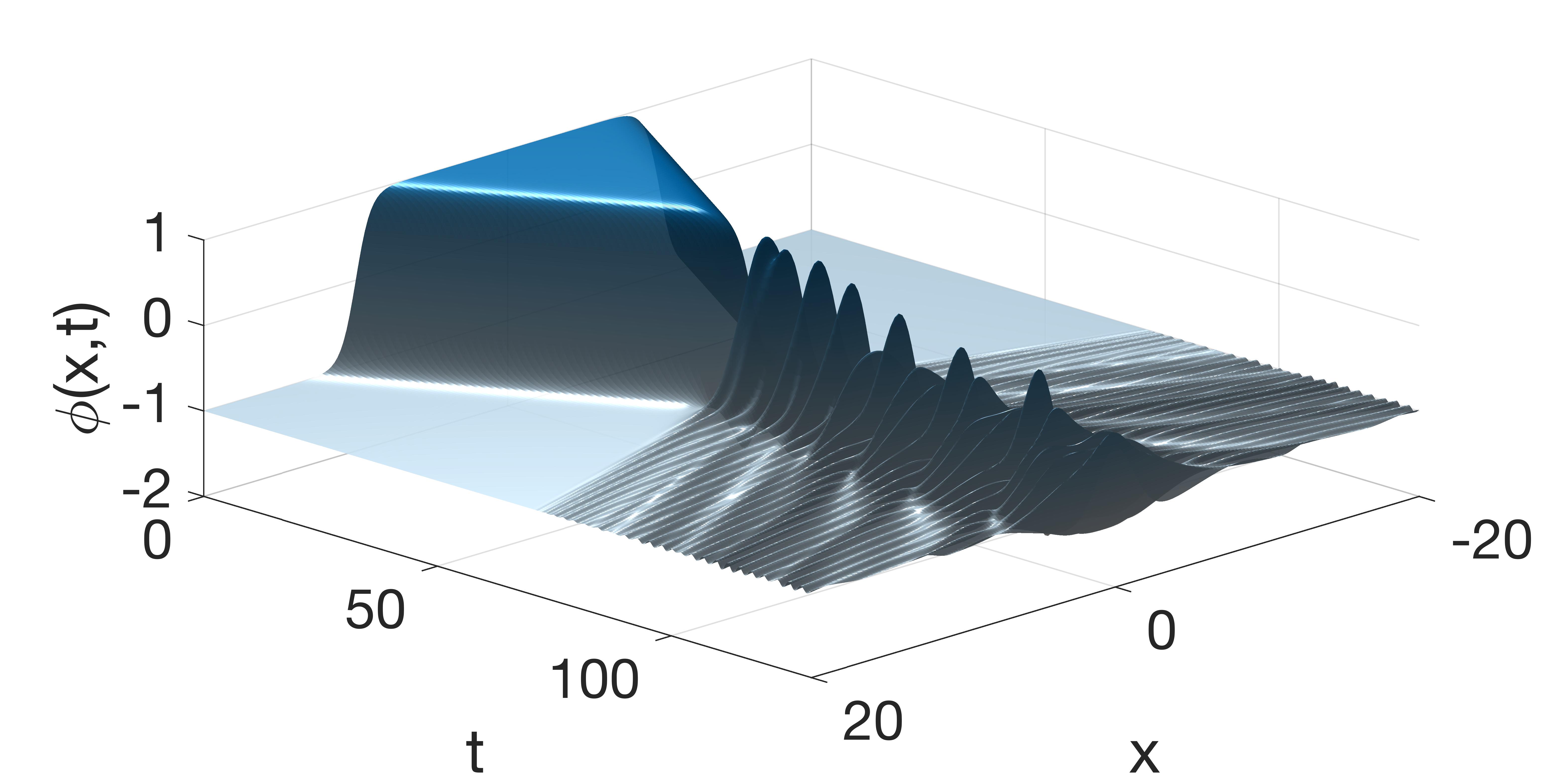}
\end{subfigure}
\hfill
\begin{subfigure}[b]{0.325\textwidth}
\centering
\includegraphics[width=\textwidth]{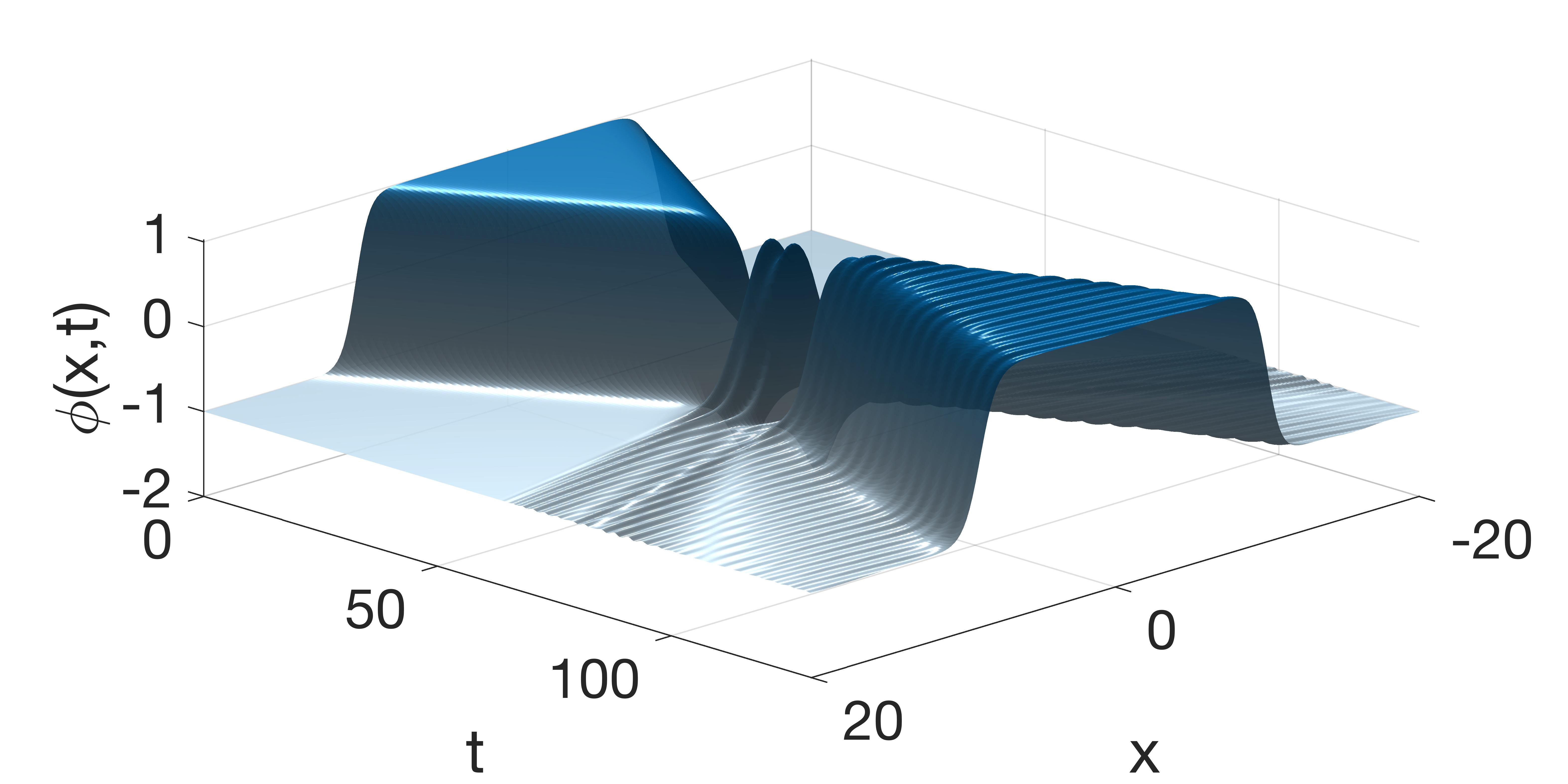}
\end{subfigure}
\hfill
\begin{subfigure}[b]{0.325\textwidth}
\centering
\includegraphics[width=\textwidth]{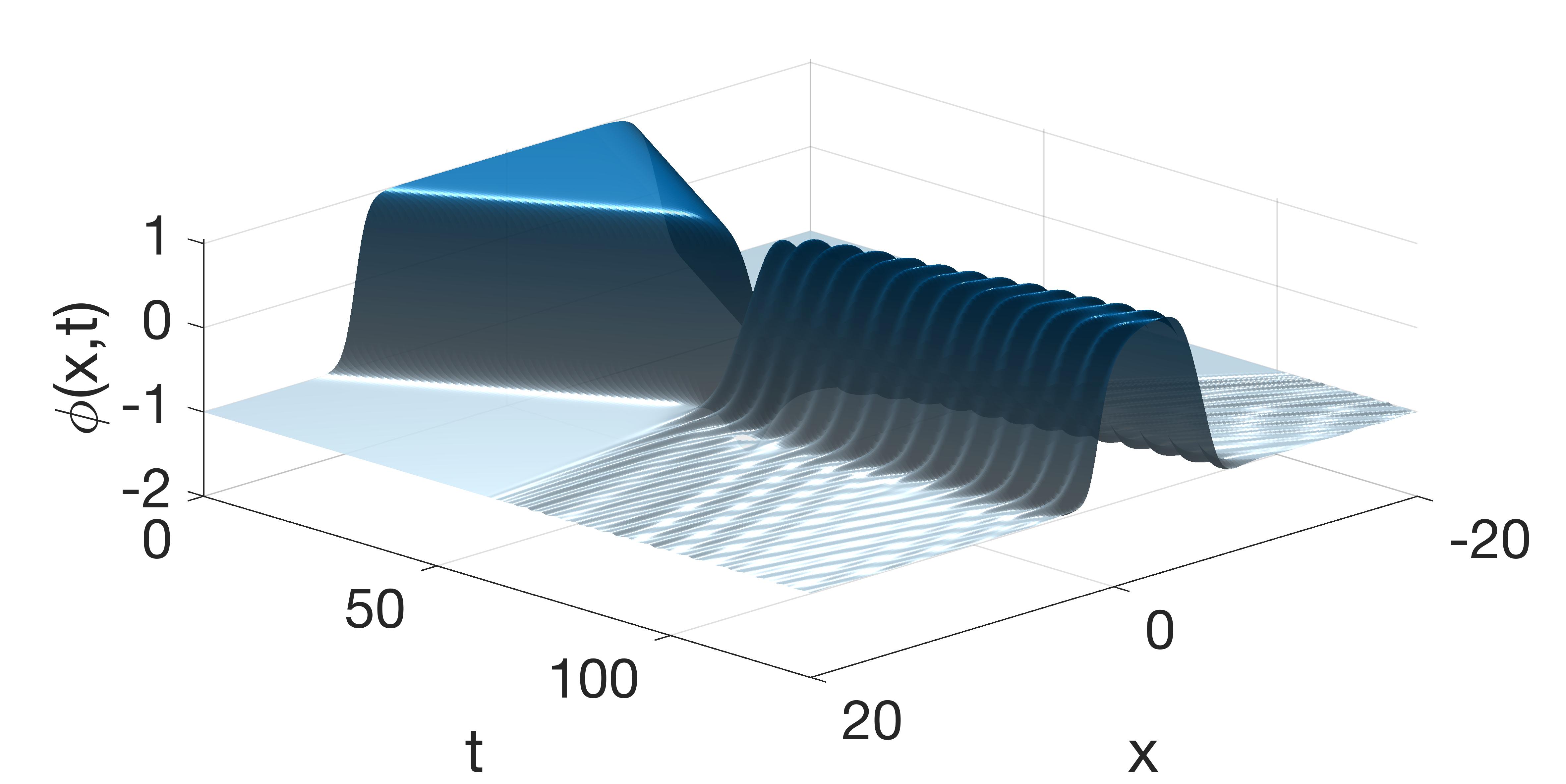}
\end{subfigure}
\quad
\begin{subfigure}[b]{0.325\textwidth}
\centering
\includegraphics[width=\textwidth]{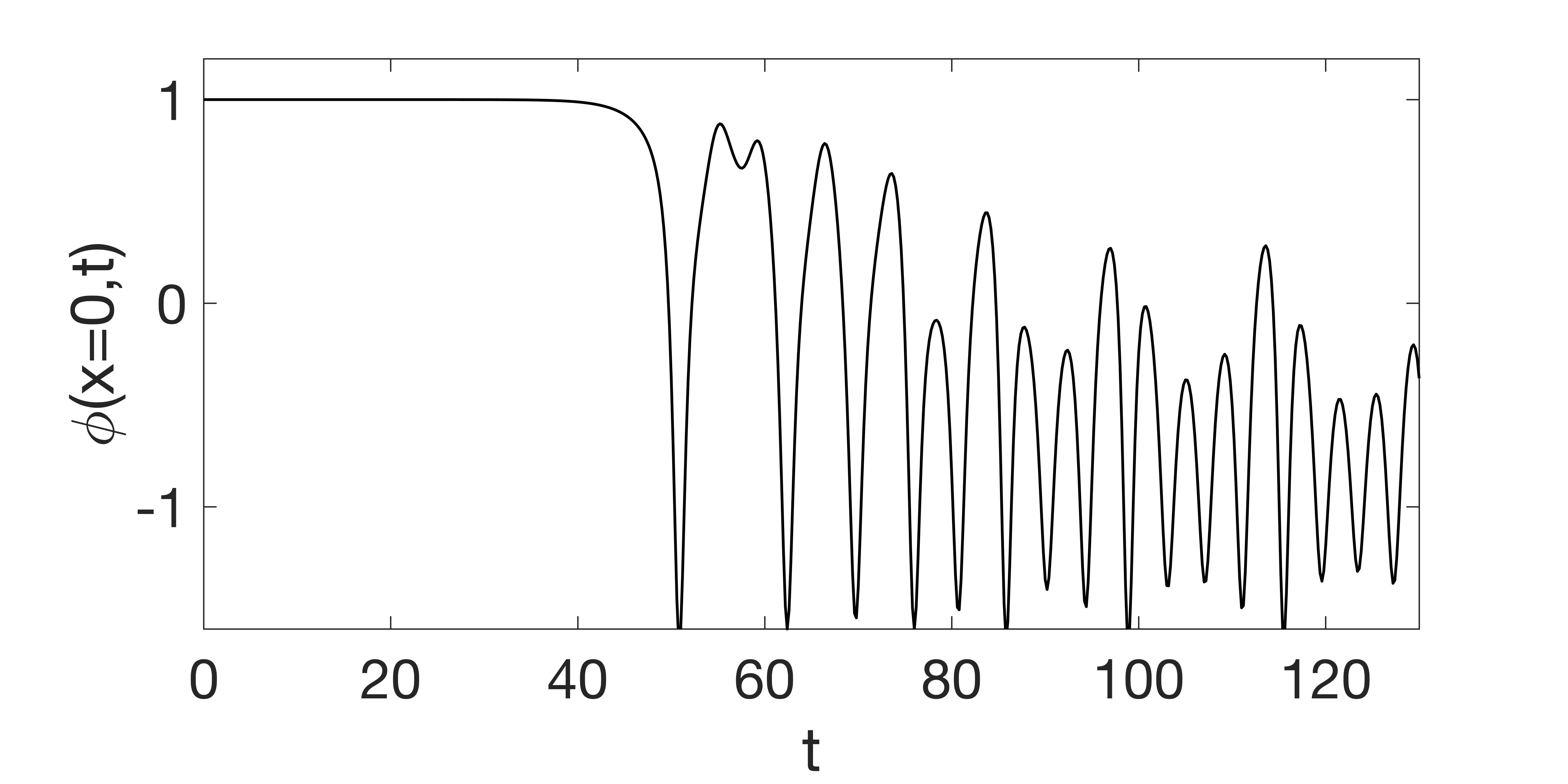}
\end{subfigure}
\hfill
\begin{subfigure}[b]{0.325\textwidth}
\centering
\includegraphics[width=\textwidth]{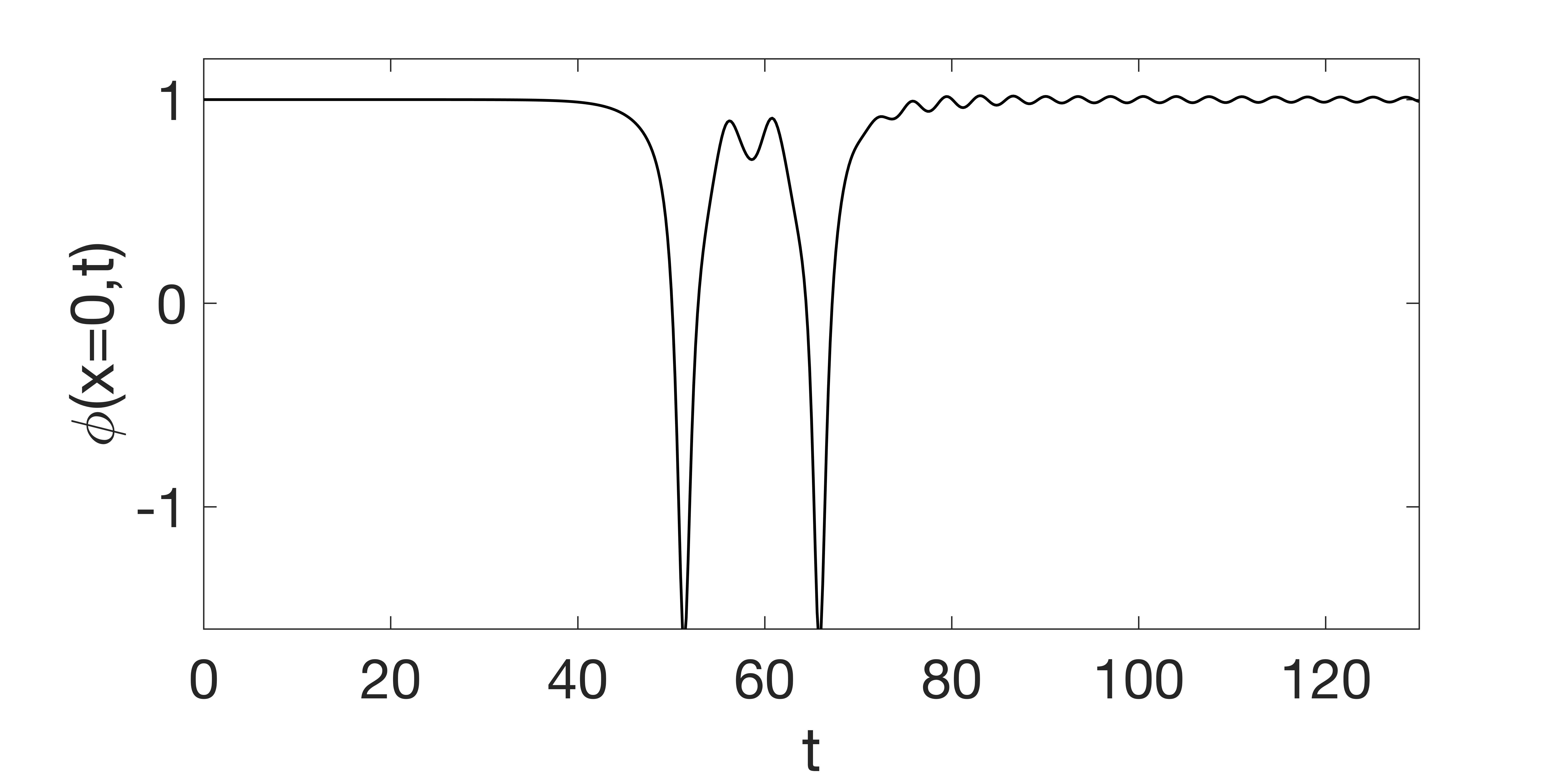}
\end{subfigure}
\hfill
\begin{subfigure}[b]{0.325\textwidth}
\centering
\includegraphics[width=\textwidth]{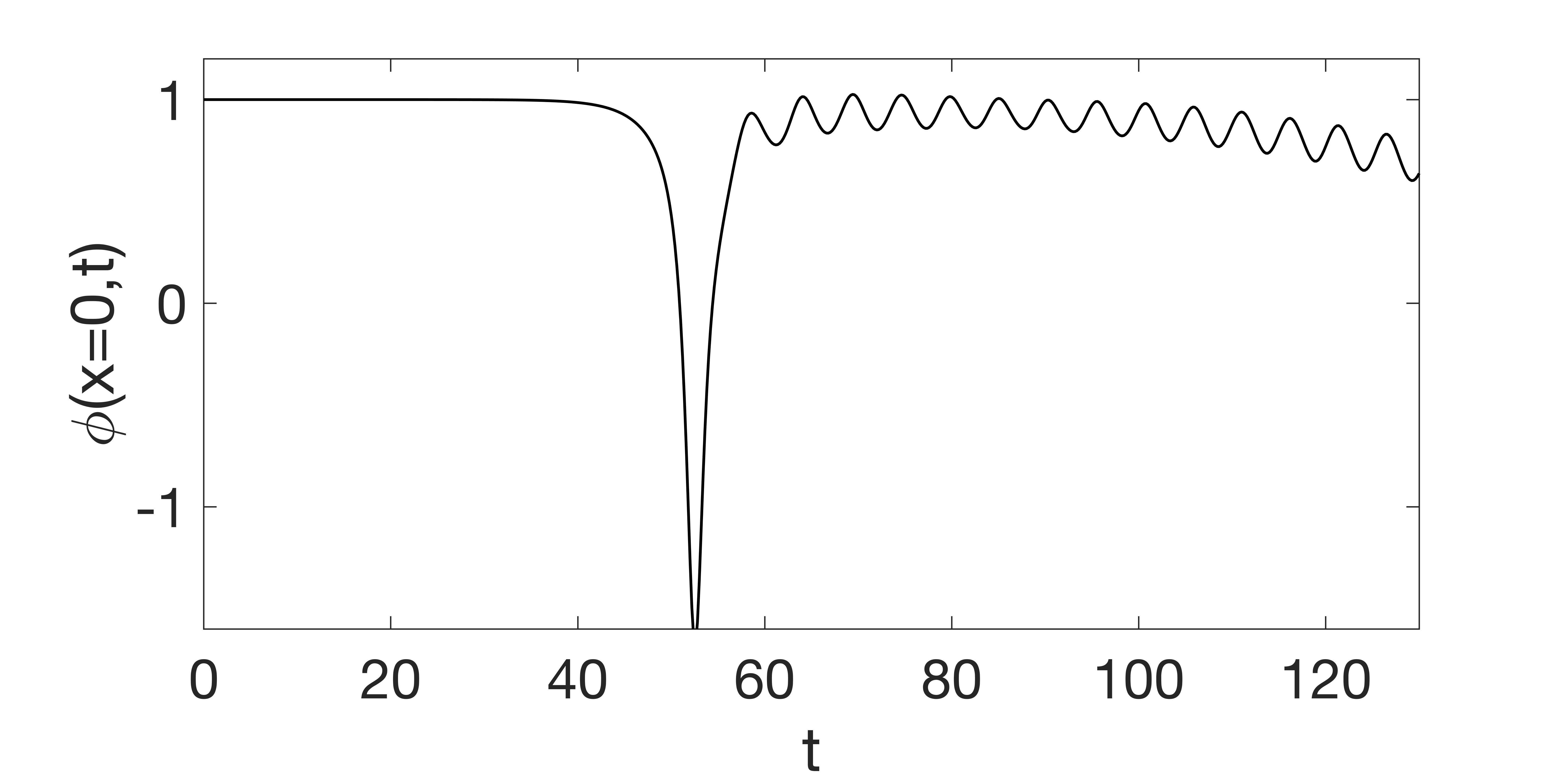}
\end{subfigure}
\quad
\begin{subfigure}[b]{0.325\textwidth}
\centering
\includegraphics[width=\textwidth]{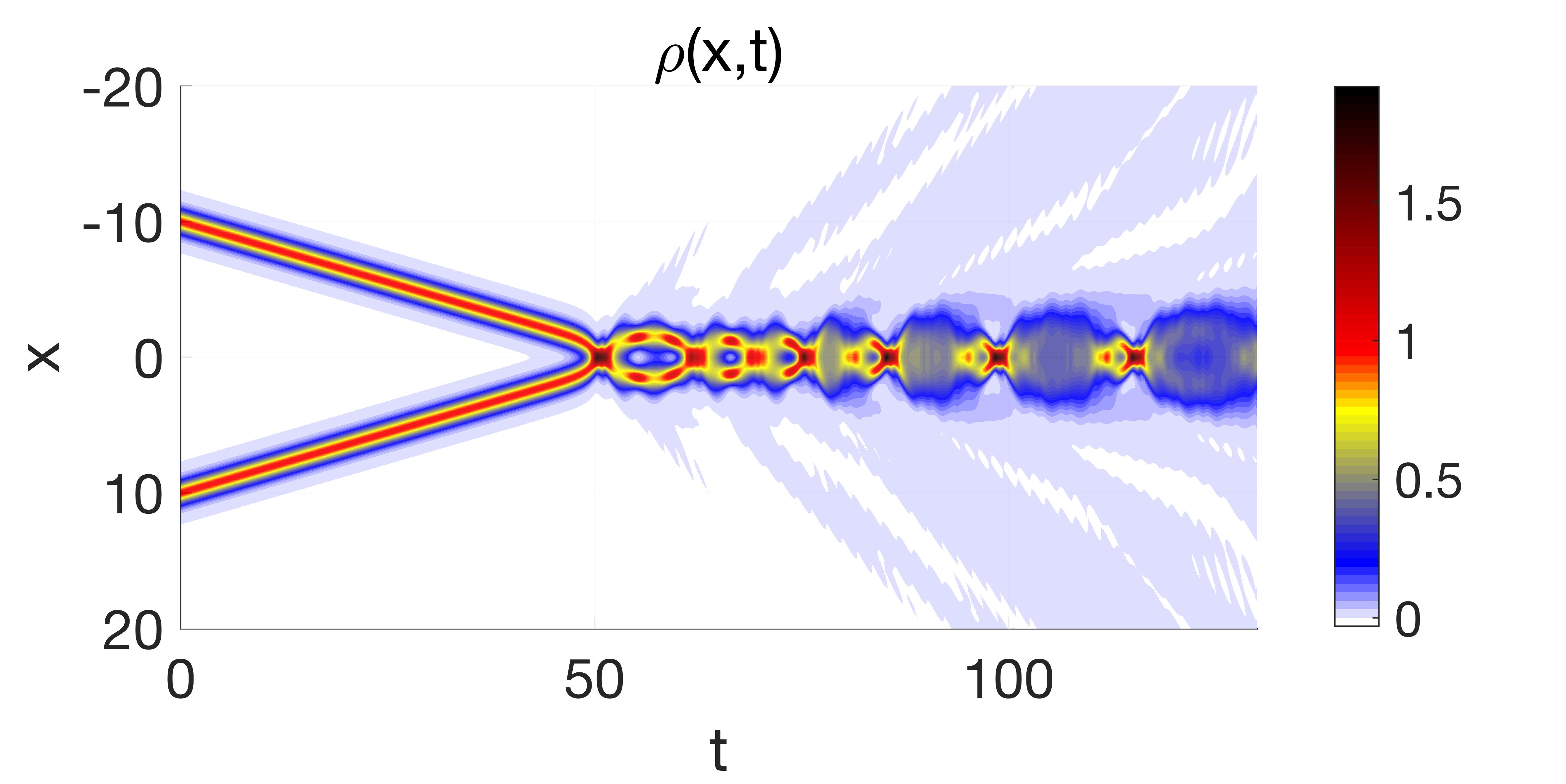}
\caption{$\alpha=0$, bion.}
\label{larger_vc}
\end{subfigure}
\hfill
\begin{subfigure}[b]{0.325\textwidth}
\centering
\includegraphics[width=\textwidth]{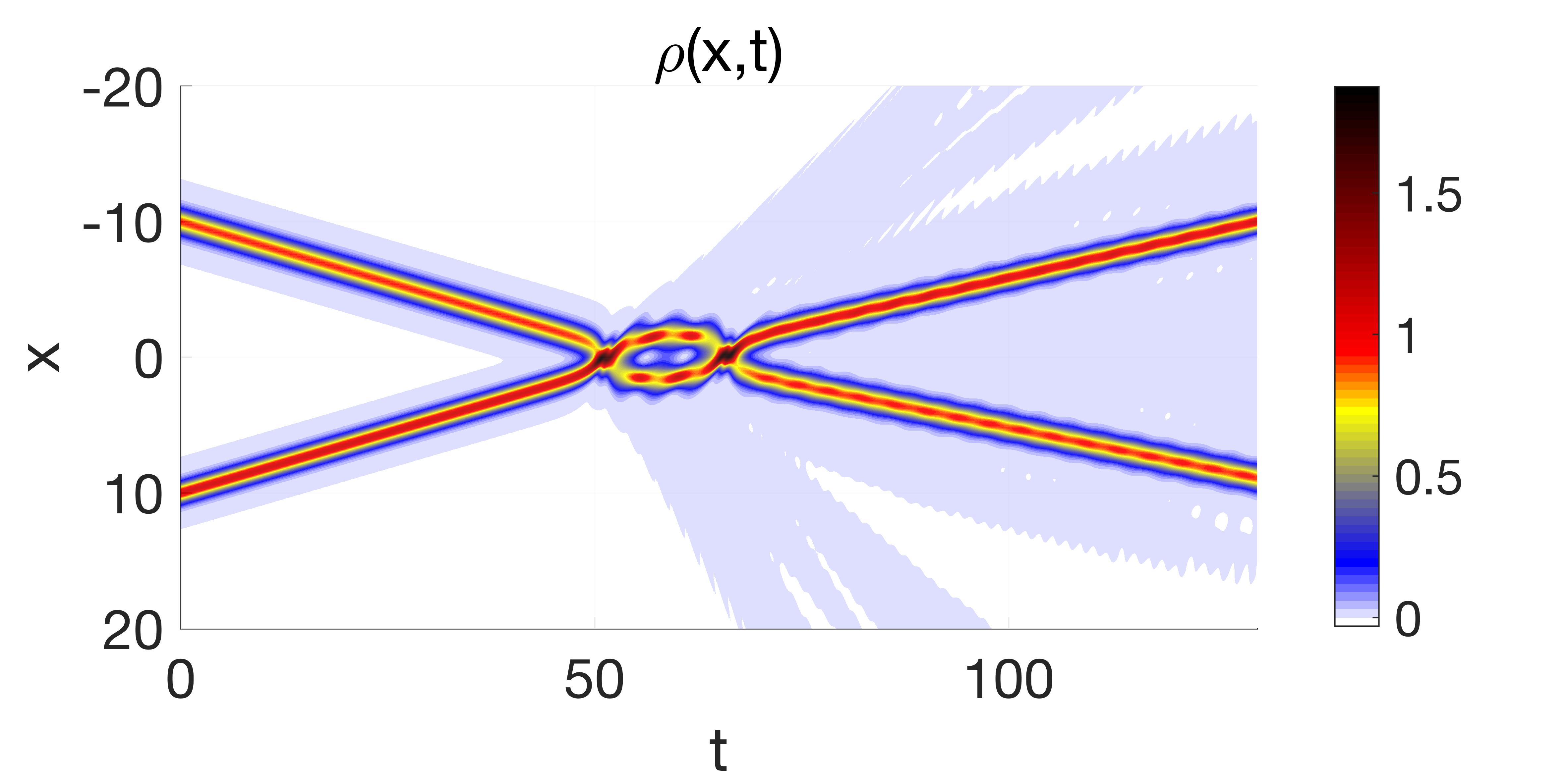}
\caption{$\alpha=0.5$, two bounce.}
\label{less_vc}
\end{subfigure}
\hfill
\begin{subfigure}[b]{0.325\textwidth}
\centering
\includegraphics[width=\textwidth]{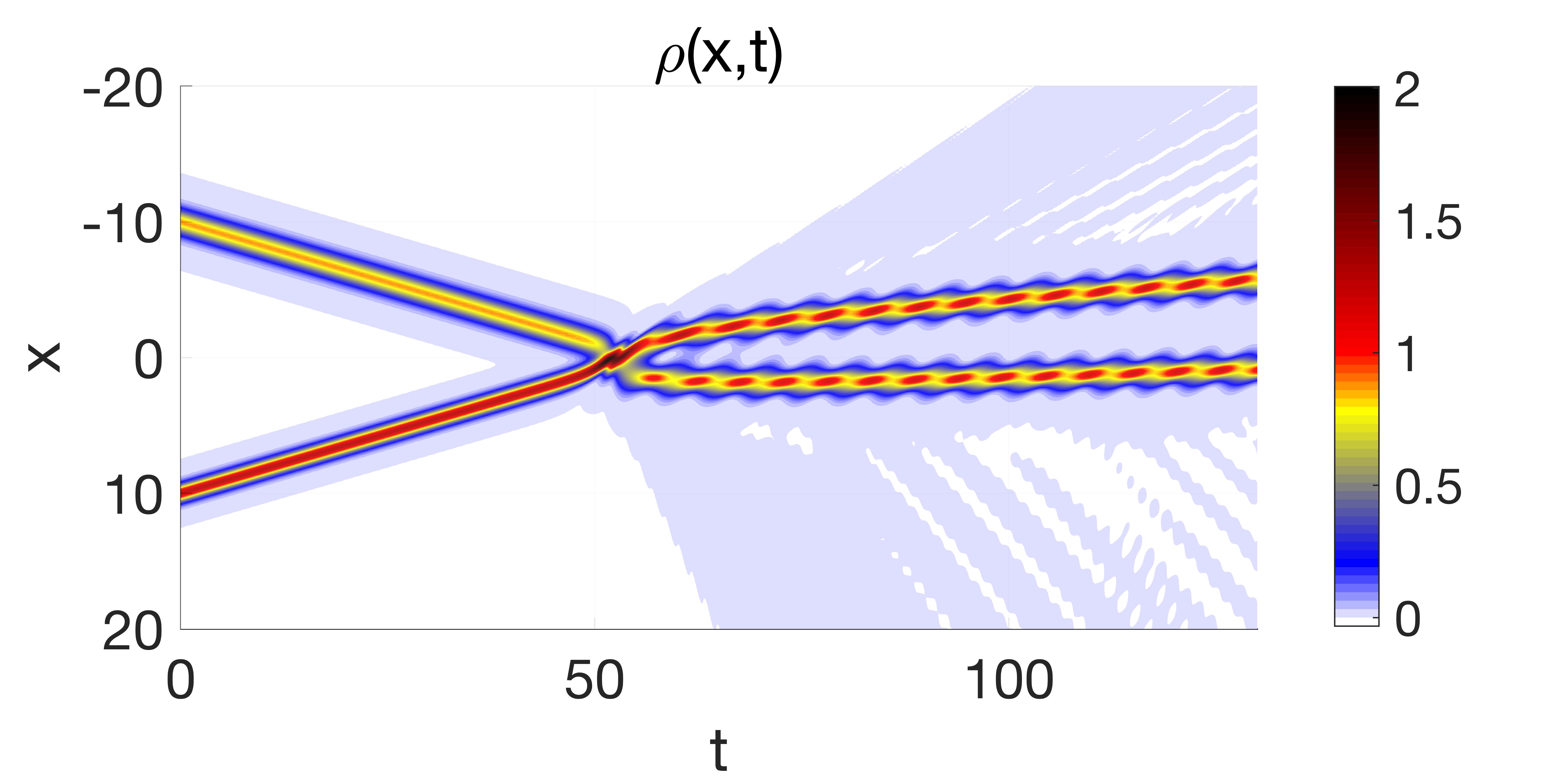}
\caption{$\alpha=1$, inelastic scattering.}
\label{tbw_collision}
\end{subfigure}
\caption{Field configuration (top row), field value at the origin (middle row), and corresponding energy density (bottom row) for $v_0=0.178$. The Lorentz-violating parameters $\alpha$ corresponding to the left-hand, middle, and right-hand columns are $0, 0.5$, and 1, respectively, which
give bion, two-bounce, and inelastic scattering solutions, respectively.
 These computations are conducted with $\Delta x=0.2$, RelTol=$10^{-9}$ and AbsTol=$10^{-10}$. The typical relative error of the total energy is $|\delta E|\lesssim10^{-9}$.}
\label{collision}
\end{figure*}

As an explicit example, the evolution of field configuration and the corresponding energy density for $v_0=0.178$ are plotted in Fig.~\ref{collision}.
With this initial velocity, one would observe bion, two-bounce, and inelastic scattering by taking $\alpha=0, 0.5$, and 1, respectively. From the bottom panels of Fig.~\ref{collision}, it can also be seen that the radiation emitted after collisions is asymmetric with respect to the origin.

\begin{figure}[h]
\centering
\includegraphics[width=0.49\textwidth]{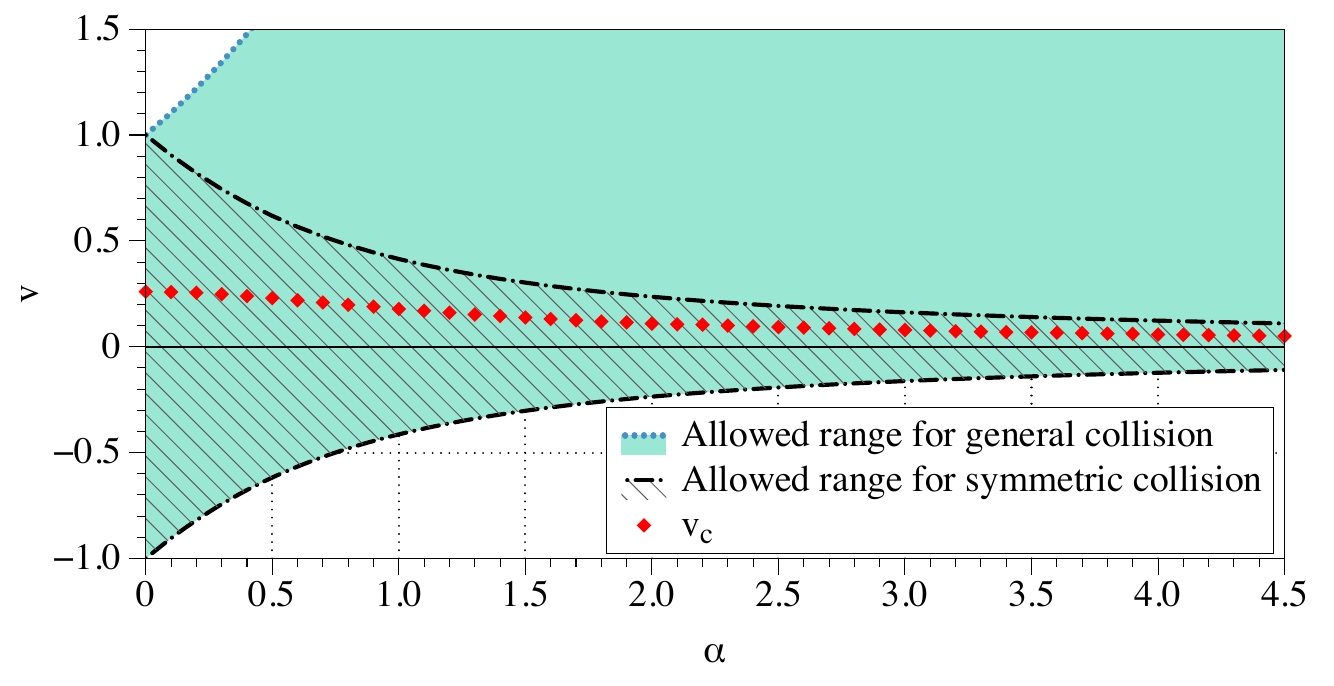}
\caption{Critical velocity $v_{c}$ (red dots) as a function of the Lorentz-violating parameter $\alpha\in[0,4.5]$. Areas filled with green and with a slashed pattern are the allowed parameter spaces for general and symmetric collisions, respectively.}
\label{fig_vc}
\end{figure}

\begin{figure}[h]
\centering
\includegraphics[width=0.47\textwidth]{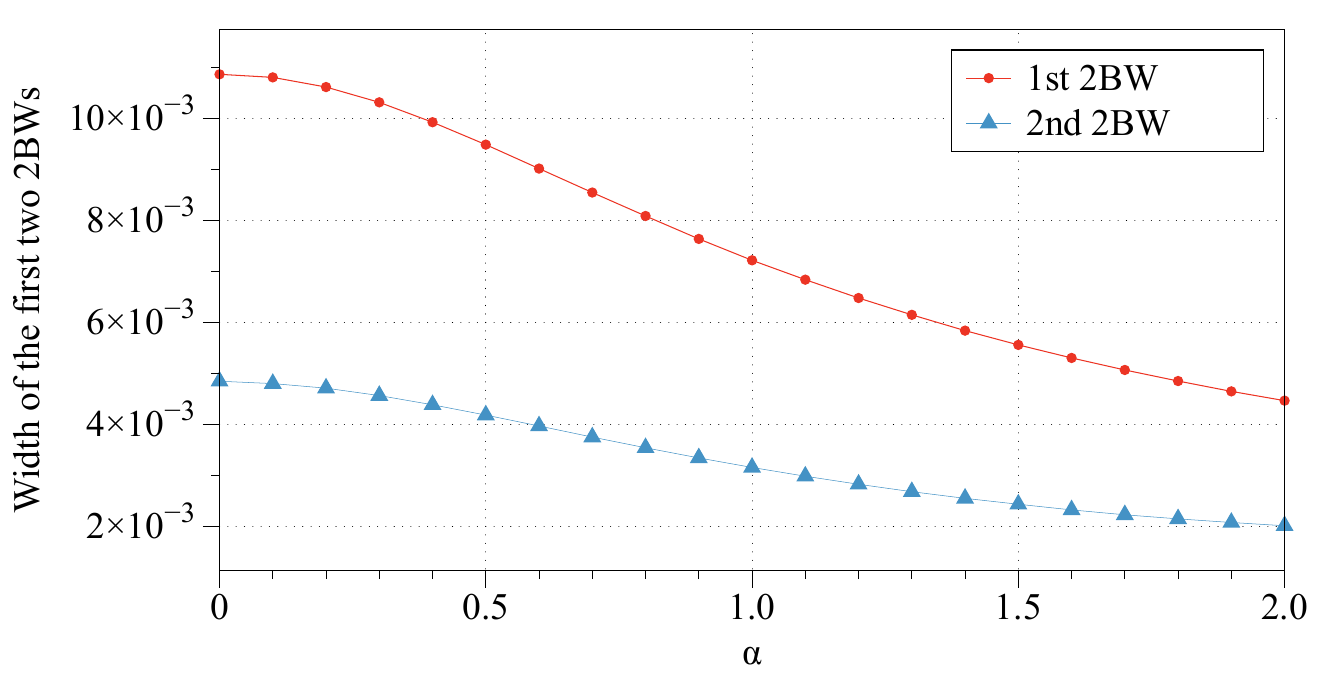}
\caption{Widths of the first two 2BWs.}
\label{bw_figure}
\end{figure}

The critical velocity $v_c$ for $\alpha\in [0,4.5]$ is plotted in Fig.~\ref{fig_vc}, which shows that $v_c$ is a monotonically decreasing function of $\alpha$, and it is always below the maximum velocity $v_{\rm 0, max}$ allowed for a velocity-symmetric collision.  Figure \ref{bw_figure} shows how the widths of the first two 2BWs vary with respect to $\alpha$. It also decreases  monotonically as $\alpha$ becomes larger.

\subsection{Information from maxima of energy densities}
Given a value of $\alpha$, the fractal structure clearly shows how the collision results vary with the initial velocity $v_0$. One may ask the reverse question; that is, given a value of $v_0$, can one tell which interval of parameter $\alpha$ corresponds to a two-bounce, three-bounce, or inelastic collision? In Ref.~\cite{GaniMarjanehSaadatmand2019}, the authors found that the maxima of various kinds of energy densities can provide important information on the collision phenomena. To see this, the energy density defined in Eq.~\eqref{rhos} is first divided as follows:
\begin{equation}
\rho(x,t)=k(x,t)+u(x,t)+p(x,t),
\end{equation}
where $k(x,t)=\frac{1}{2}\left(\frac{\partial \phi}{\partial t}\right)^{2}$, $u(x,t)=\frac{1}{2}\left(\frac{\partial \phi}{\partial x}\right)^{2}$,
and $p(x,t)=V(\phi)$ are the kinetic energy density, elastic strain energy, and on-site potential energy, respectively (also see~\cite{SaadatmandDmitrievKevrekidis2015,MarjanehSaadatmandZhouDmitrievEtAl2017}).

\begin{figure}[h]
\centering
\includegraphics[width=0.48\textwidth]{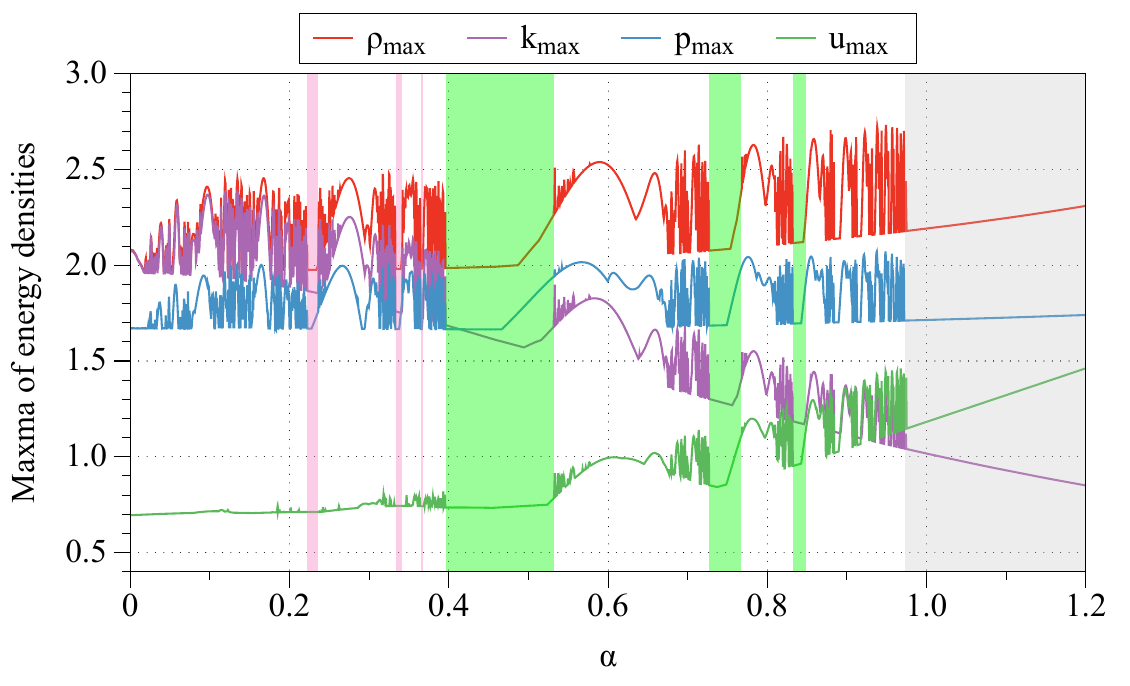}
\caption{(Color online) Maxima of energy densities for velocity-symmetric kink-antikink collisions with $v_0 = 0.178$ and $\alpha\in[0, 1.2]$. The first three 2BWs are highlighted (green zones), as are the first three 3BWs around the first 2BW (pink zones). The gray zone corresponds to inelastic scattering, and unhighlighted chaotic segments correspond to higher-order bounce windows and bions.}
\label{max_dens_plot}
\end{figure}

\begin{figure}[h]
\centering
\includegraphics[width=0.48\textwidth]{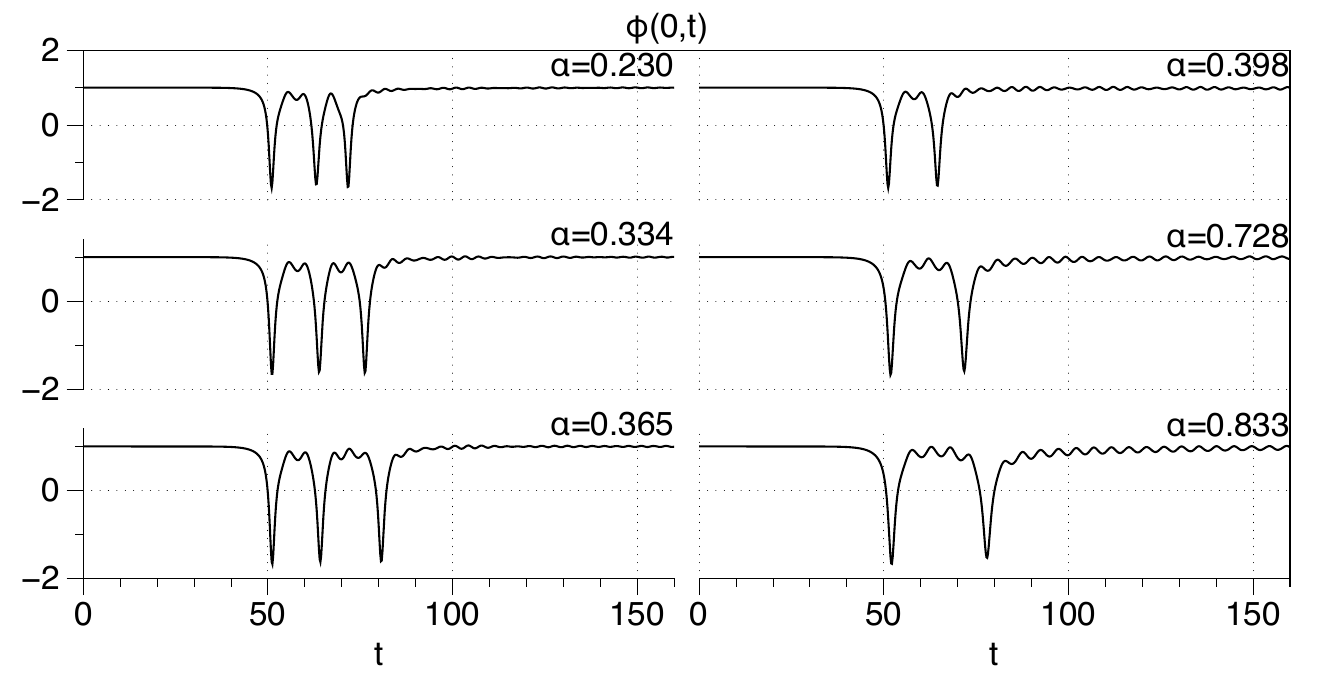}
\caption{Curves of $\phi(x=0,t)$ of some representative values of $\alpha$ in 3BWs and 2BWs highlighted in Fig.~\ref{max_dens_plot}. Initial velocity is taken as $v_0=0.178$. }
\label{fig_maxEnergy_rho}
\end{figure}

By simulating velocity-symmetric kink-antikink collisions with $v_0=0.178$ and $\alpha\in [0, 1.2]$, curves of the maxima of the aforementioned four types of energy densities are obtained: $\rho_{\rm max}, k_{\rm max}, u_{\rm max}$, and $p_{\rm max}$ (see Fig.~\ref{max_dens_plot}). These curves are qualitatively similar; that is,
they show a fractal structure, i.e.,
in some intervals they behave in an orderly manner, but in several others they seem to be chaotic. The ordered intervals correspond to $n$-bounce windows. The first three 2BWs, which  correspond to $\alpha\in[0.3975,0.5310]$,  $[0.7275,0.7670]$, and $[0.8330,0.8515]$, are highlighted in green, and
the first three 3BWs around the first 2BW, which correspond to $\alpha\in[0.2225,0.2360]$, $[0.3335,0.3415]$, and $[0.3650,0.3680]$, in pink.
In Fig. \ref{fig_maxEnergy_rho}, the $\phi(0,t)$ curves for the two- and three-bounce solutions are plotted, which
correspond to some representative values of $\alpha$ that lie in the highlighted bounce windows.

To our knowledge, this is the first report on the fractal structure in the maximal energy density graph, despite the fact that the maximal density graph has been used in many other aspects~\cite{AskariSaadatmandDmitrievJavidan2018,GaniMarjanehSaadatmand2019}.

\section{Summary}
\label{sec_con}

In this work, kink-antikink collisions of a Lorentz-violating $\phi^4$ model have been studied. After a short summary of the model and the kink solution of Ref.~\cite{BazeiaMenezes2006}, the linear
perturbation spectrum of the static kink solution was analyzed. It was found that the Lorentz-violating term of the proposed model does not change the spectrum structure of the standard $\phi^4$ model, so the static kink solution is linearly stable and there exists only one vibrational mode apart from a zero mode. However the Lorentz-violating term does impact the wave function and frequency of the vibrational mode. As a consequence, the kink-antikink collision phenomena of the present model will deviate from those of the standard $\phi^4$ model.

The deviation was studied numerically via two different approaches. In the first approach,
fixing the values of the Lorentz-violating parameter to be $\alpha=0, 0.5, 1$, the collision with
the initial velocity $v_0$ ranging from 0.12 to 0.27 was scanned. By comparing the fractal structures of each value of $\alpha$ (Fig.~\ref{structure}), it was found that models with larger Lorentz-violating parameters have smaller critical velocities and narrower widths of 2BWs. The values of critical velocity and
widths of 2BWs were also calculated for arbitrary values of $\alpha$ (see Figs.~\ref{fig_vc} and \ref{bw_figure}).

In the second approach, setting the initial velocity as $v_0=0.178$, the maximal energy densities corresponding to $\alpha\in[0,1.2]$ were scanned. An interesting fractal structure was observed for the first time (see Fig.~\ref{max_dens_plot}). In the curves of maximal energy densities, the intervals corresponding to bions are more chaotic than those of the two bounces, three bounces, and inelastic scatterings. This indicates that the plot of maximal energy densities might be very useful in analyzing the results of kink collisions.

Phenomenological applications of the present work are worth consideration, but go beyond the scope of the present work.

\section*{Acknowledgements}

Yuan Zhong thanks Zhi Xiao for discussions on Lorentz violation. This work was supported by the National Natural Science Foundation of China (Grant Nos.~11847211, 11605127 and 11875151), Fundamental Research Funds for the Central Universities (Grant No.~xzy012019052), and China Postdoctoral Science Foundation (Grant No.~2016M592770). It was also supported in part by JSPS KAKENHI Grant Nos. JP17H06359 and JP19K03857, and by Waseda University Grant for Special Research Projects (Project Nos. 2019C-254 and 2020C-270).

\section*{Bibliography}



\end{document}